\documentclass[11pt]{revtex4-1}

\usepackage{graphicx}

\begin{document}

\title{Thermodynamics of the inhomogeneous perfect fluid LTB model: Modified Bekenstein-Hawking system}

\author{Subhajit Saha\footnote {subhajit1729@gmail.com}}
\author{Subenoy Chakraborty\footnote {schakraborty.math@gmail.com}}

\affiliation{Department of Mathematics, Jadavpur University, Kolkata 700032, West Bengal, India.}

\begin{abstract}

The present work deals with three alternative generalized Bekenstein-Hawking formulation of thermodynamical parameters namely entropy and temperature for the universal thermodynamical system bounded by a horizon in the frame work of inhomogeneous perfect fluid Lemaitre-Tolman-Bondi (LTB) model of the Universe. 
For the first choice, the first law of thermodynamics holds only for the trivial de Sitter case of the LTB model while we need restriction on the evolution of the horizon radius for the validity of the generalized second law of thermodynamics. However, for the other two choices, the first law of thermodynamics holds with some specific (integral) form of the parameters involved but for generalized LTB models.

\end{abstract}

\maketitle

\section{INTRODUCTION}

In 1975, Hawking \cite{Hawking1} was able to show emission of radiation from black holes using semi-classical description. Since then, black hole (BH) is considered as a thermodynamical object, i.e., laws of BH physics and thermodynamical laws are equivalent \cite{Bekenstein1, Bardeen1}. Subsequently, there have been a lot of work dealing with thermodynamical studies of the Universe bounded by an apparent horizon \cite{Jacobson1, Padmanabhan1, Cai1, Akbar1, Cai2, Bousso1, Cai3, Wang1} or by an event horizon \cite{Mazumder1, Dutta1, Chakraborty1, Chakraborty2}, specially for the homogeneous and isotropic FRW model of the Universe. Also, for the inhomogeneous LTB model \cite{Chakraborty3}, Einstein field equations were shown to be equivalent with the unified first law of thermodynamics (UFLT) on the apparent horizon and the generalized second law of thermodynamics (GSLT) has been shown to be valid both on the apparent as well as on the event horizon with certain restrictions.

In 2006, Wang et al. \cite{Wang1}, based on a comparative study of the thermodynamical laws in the FRW model of the Universe bounded by apparent and event horizons, concluded that the Universe bounded by an apparent horizon is a perfect thermodynamical system (Bekenstein system), while both the first and second laws of thermodynamics break down on an event horizon, i.e., an unphysical system. Very recently, a generalized Hawking temperature (which coincides with the Hawking temperature on the apparent horizon) or a modified Bekenstein entropy has been introduced \cite{Chakraborty4} on the event horizon and it has been possible to show the validity of both the thermodynamical laws on the event horizon irrespective of any fluid distribution in the FRW model. In the present work, we make an attempt to extend the idea of generalized Hawking temperature or modified Bekenstein entropy for the inhomogeneous LTB model and examine the validity of both the thermodynamical laws on an arbitrary horizon. Current supernova and some other data \cite{Moffat1} provides a justification in support of the inhomogeneous LTB model of the Universe. Also, using perturbation analysis, Clarkson and Maartens \cite{Clarkson1} have given a justification for the inhomogeneous model. It should be noted that Refs. \cite{Moffat1} and \cite{Clarkson1} consider only LTB models with a dust source in the context of fitting observational data. However, in the present context, LTB models with a perfect fluid source may not be useful to fit observations unless it is possible to have a physical interpretation of a dark energy and dark matter mixture as in Ref. \cite{Sussman1}.

\section{BASIC EQUATIONS IN THE INHOMOGENEOUS LTB MODEL}

The line element for the inhomogeneous spherically symmetric Lemaitre-Tolman-Bondi (LTB) spacetime in a comoving frame is given by
\begin{equation}
ds^2=h_{ab}dx^adx^b+R^2d{\Omega _2}^2
\end{equation}  
where $a$, $b$ can take values 0 and 1, and the two-dimensional metric tensor $h_{ab}$ (known as normal metric) is given by
\begin{equation}
h_{ab}=\text{diag}\left(-1,\frac{R'^2}{1+f(r)}\right)
\end{equation}
with $x^a$ being the associated co-ordinates ($x^0=t$, $x^1=r$). Here $R=R(r,t)$ is the areal radius of the spherical surface (a scalar field in the normal 2D space) and the curvature scalar $f(r)$ classifies the spacetime as bounded, marginally bounded, or unbounded according as $-1<f(r)<0$, $f(r)=0$, or $f(r)>0$.

Now by introducing the mass function $F(r,t)$ (related to the mass within the comoving radius $r$) \cite{Joshi1} as
\begin{equation}
F(r,t)=R({\dot{R}}^2-f(r)),
\end{equation}
the Einstein field equations read ($G=1=c$)
\begin{eqnarray} \label{efeltb}
8\pi \rho=\frac{F'(r,t)}{R^2R'}~~~~~~~~\text{and}~~~~~~~~8\pi p=-\frac{{\dot{F}}(r,t)}{R^2\dot{R}}
\end{eqnarray}
and the evolution equation for $R$ is given by
\begin{eqnarray}
2R\ddot{R}+{\dot{R}}^2+8\pi pR^2=f(r).
\end{eqnarray}
In the above, $\rho$ and $p$ are the energy density and thermodynamic pressure corresponding to perfect fluid having energy-momentum tensor
\begin{eqnarray}
T_{\mu \nu}=(\rho +p)u_{\mu} u_{\nu} +pg_{\mu \nu},
\end{eqnarray}
where the fluid 4-velocity $u^{\mu}$ is normalized by $u_{\mu} u^{\nu}=-1$. The energy-momentum conservation relation ${T_\mu}^\nu_{;\nu}=0$ for the above spacetime model results
\begin{equation}
\dot{\rho}+3H(\rho +p)=0~,~~~p'=0,
\end{equation}
with $H=\frac{1}{3}\left(\frac{{\dot{R}}'}{R'}+2\frac{\dot{R}}{R}\right)$ as the Hubble parameter. It should be noted that perfect fluid LTB models are very restrictive as their physical interpretation is difficult because $p=p(t)$ while $\rho =\rho (t,r)$ (see Sec. 2.13 of Ref. \cite{Krasinski1}). However, a reasonable physical interpretation of this fluid source can be given \cite{Sussman1} as an interactive mixture of two fluids --- inhomogeneous dust (taken as cold dark matter) and a homogeneous dark energy fluid. On the other hand, in contrast to perfect fluid sources with an LTB metric, dust sources with this metric (the usual LTB models) are physically well motivated. Special cases of density voids constructed with these models were shown \cite{Mishra1} to be inconsistent with the Bekenstein-Hawking formalism. In this context, we shall now examine whether perfect fluid LTB model will be a Bekenstein-Hawking thermodynamical system or not.

We now introduce another relevant scalar quantity on this normal space as
\begin{eqnarray}
\chi (x)=h^{ab}{\partial _a}R{\partial_ b}R=1-\frac{F(r,t)}{R}.
\end{eqnarray}
Usually, the apparent horizon is defined at the vanishing of this scalar, {\it i.e,} $\chi (x)=0$, which gives
\begin{equation}
R_A=F(r,t).
\end{equation}
Surface gravity is an important quantity in classical general relativity (GR) which plays a vital role in black hole (BH) thermodynamics and semi-classical aspects of gravity, being closely related to the temperature of Hawking radiation. For a static or stationary BH, the surface gravity on the BH is a constant (the zeroth law of BH thermodynamics). For a dynamic horizon, the surface gravity not only depends on the horizon radius but also on the time derivative of the horizon radius. Usually the surface gravity at the horizon is defined as 
\begin{equation}
\kappa _h=\frac{1}{2\sqrt{-h}}\partial _a \left(\sqrt{-h} h^{ab} \partial _b R \right).
\end{equation}
For the present LTB model, it can be written in three equivalent forms as
\begin{equation}
\kappa _h=\left\{
\begin{array}{lll}
\frac{1}{2}\left(-\frac{\dot{R}'\dot{R}}{R'}-\ddot{R}+\frac{f'(r)}{2R'}\right)_{R=R_h} \\
\left\lbrace -\frac{1}{4R}\left(\frac{\dot{F}}{\dot{R}}+\frac{F'}{R'}-\frac{2F}{R}\right)\right\rbrace _{R=R_h} \\
\left[ -\frac{1}{4R} \left\lbrace 8\pi R^2 (\rho -p)-\frac{2F}{R}\right\rbrace \right]_{R=R_h}
\end{array}
\right. .
\end{equation}
and hence we have the Hawking temperature on the horizon as
\begin{equation}
T_h=\frac{|\kappa _h|}{2\pi}.
\end{equation}

\section{VALIDITY OF THE FIRST AND THE SECOND LAWS OF THERMODYNAMICS}

To evaluate the energy flow across the horizon, we consider UFLT on the horizon \cite{Chakraborty3, Cai4}, i.e., 
\begin{equation}
dE_h=A\Psi +WdV,
\end{equation}
where $A=4\pi R_{h}^{2}$, $V=\frac{4}{3}\pi R_{h}^{3}$ are respectively the area and volume bounded by the horizon, $\Psi=\psi _adx^a$, $\psi _a=T_{a}^{b}{\partial _b}R+W{\partial _a}R$ is the energy flux (or momentum density) and $W=-\frac{1}{2}\text{Trace}(T)$ is the work function, with trace over the 2D space normal to the spheres of symmetry. For the present model, $\Psi =-\frac{1}{2}(\rho +p)(\dot{R}dt-R'dr)$,~$W=\frac{1}{2}(\rho -p)$ and hence
\begin{equation}
dE_h=\left\lbrace 4\pi R^2(\rho R'dr-p\dot{R}dt)\right\rbrace _{R=R_h}=\frac{1}{2}dF_h,
\end{equation}
where the last equality is obtained by using the Einstein field equations (\ref{efeltb}).

Now, if we use the Bekenstein's entropy-area relation on the given horizon and the temperature on the horizon [known as the generalized Hawking temperature (GHT)] as \cite{Chakraborty4}
\begin{equation}
T_{h}^{(g)}=\alpha {T_h},
\end{equation}
where the (positive) dimensionless parameter $\alpha$ is chosen such that the first law of thermodynamics (FLT) holds on the horizon. Thus, in order to satisfy the Clausius relation (i.e., the FLT) $dE_h=T_{h}^{(g)} dS_h$, we have (at $R=R_h$)
\begin{equation} \label{cond1}
\frac{F'}{R'}=-\frac{\alpha}{2} \left[8\pi R^2 (\rho -p)-\frac{2F}{R}\right]
\end{equation} 
and
\begin{equation} \label{cond2}
\frac{\dot{F}}{\dot{R}}=-\frac{\alpha}{2} \left[8\pi R^2 (\rho -p)-\frac{2F}{R}\right]
\end{equation}
Hence for the validity of the FLT, we must have
\begin{equation} \label{frd-fr}
\frac{F'}{R'}=\frac{\dot{F}}{\dot{R}},
\end{equation}
which is the integrability condition for the general one-form
\begin{equation} \label{dom}
d\Omega =RdF.
\end{equation}
Here, $\Omega$ is a smooth function of $r$ and $t$. Now, in order to determine $\Omega$, from Eq. (\ref{dom}), we have---
\begin{equation}
\dot{\Omega}=R\dot{F}~~~~~~~~\text{and}~~~~~~~~\Omega '=RF'.
\end{equation}
Using Eqs. (\ref{cond1}) and (\ref{cond2}) along with the Einstein field equations (\ref{efeltb}), we obtain
\begin{equation} \label{rho-p}
\rho +p=0,
\end{equation}
for any $\Omega$.
Hence for the validity of the FLT (for any choice of $\alpha$), the horizon should satisfy the de Sitter condition $\rho +p=0$, i.e., the de Sitter particular case of the model. In other words, the FLT is not valid for non-trivial LTB models.

Next, in order to examine the validity of the GSLT, we start with the Gibb's law \cite{Wang1, Izquierdo1} to determine the entropy variation of the bounded fluid distribution, i.e.,
\begin{equation}
T_{fl}dS_{fl}=dE_{fl}+pdV_h
\end{equation}
where $T_{fl}$ and $S_{fl}$ are respectively the temperature and entropy of the given fluid distribution, $V_h=\frac{4}{3}\pi R_{h}^{3}$ and $E_{fl}=\rho V_h$. The above equation explicitly takes the form
\begin{equation}
T_{fl}dS_{fl}=4\pi R_{h}^{3}(\rho +p)\left[\left\lbrace \frac{\dot{R}_h}{R_h}-H\right\rbrace dt+\left\lbrace \frac{R_{h}^{'}}{R_h}-\frac{1}{3}\frac{w'}{w(1+w)}\right\rbrace dr \right],
\end{equation}
where $w(r,t)=\frac{p(t)}{\rho (r,t)}$ can be interpreted as the equation of state parameter of the barotropic fluid. Also using the FLT $T_{h}^{(g)} dS_h=dE_h$, where $dE_h$ is given by Eq. (14), we have,
\begin{equation}
T_{h}^{(g)}dS_h=4\pi R_{h}^{3}\rho \left[-w\frac{\dot{R}_h}{R_h}dt+\frac{R_{h}^{'}}{R_h}dr\right]
\end{equation}
Now assuming $T_{fl}=T_{h}^{(g)}$, i.e., the inside fluid has the same temperature as the bounding surface, we obtain
\begin{equation} \label{gge}
T_hdS_T=4\pi R_{h}^{3}\rho \left[\left\lbrace \frac{\dot{R}_h}{R_h}-(1+w)H\right\rbrace dt+\left\lbrace (2+w)\frac{R_{h}^{'}}{R_h}-\frac{1}{3}\frac{w^{'}}{w}\right\rbrace dr\right]
\end{equation} 
with $S_T=S_{fl}+S_h$, the total entropy of the universal system. The integrability condition of the generalized Gibb's equation [Eq. (\ref{gge})] has been examined in Appendix A. For the validity of the GSLT, we must have
\begin{equation}
\frac{\dot{R}_h}{R_h} \geq (1+w)H
\end{equation}
and 
\begin{equation}
(2+w)\frac{R_{h}^{'}}{R_h} ~\geq ~\text{or}~ \leq~ \frac{1}{3}\frac{w'}{w}, ~~\text{according as} ~~dr ~> ~\text{or}~ <~ 0.
\end{equation}
However, using condition (\ref{rho-p}) for the FLT, the above restrictions for the GSLT become
\begin{equation}
\dot{R}_h \geq 0
\end{equation}
and
\begin{equation}
\dot{R}_h ~\geq ~\text{or}~ \leq~ 0 ~~\text{according as} ~~dr ~> ~\text{or}~ <~ 0.
\end{equation}
The above two conditions simplify to
\begin{equation}
dR_h ~\geq ~\text{or}~ \leq~ 0 ~~\text{according as} ~~dr ~> ~\text{or}~ <~ 0.
\end{equation}
Thus for the validity of the GSLT, the horizon radius must be an increasing function of both $r$ and $t$, and the present model is restricted to the de Sitter particular case.

\section{MODIFICATION OF BEKENSTEIN ENTROPY AND HAWKING TEMPERATURE, AND THE FLT}

Let us now examine the following two possible modifications of the Bekenstein entropy and the Hawking temperature on the horizon so that the Clausius relation holds on the horizon---
\begin{equation} \label{mod1}
\left.
\begin{array}{rcl}
S_{h}^{(m)} & = & \beta S_{h}, \\
T_{h}^{(m)} & = & T_{h},
\end{array}
\right\}
\end{equation}
and
\begin{equation} \label{mod2}
\left.
\begin{array}{rcl}
S_{h}^{(m)} & = & \delta S_{h}, \\
T_{h}^{(m)} & = & \frac{1}{\delta}T_{h},
\end{array}
\right\}
\end{equation}
where as before $\beta$ [in Eq. (\ref{mod1})] and $\delta$ [in Eq. (\ref{mod2})] are dimensionless parameters and ($S_{h}^{(m)}$, $T_{h}^{(m)}$) denote the modified Bekenstein entropy and the modified Hawking temperature on the horizon. Then for the choice in Eq. (\ref{mod1}), in order to satisfy the Clausius relation (i.e., ~$dE_h=T_{h}^{(m)}dS_{h}^{(m)}$), $\beta$ turns out to be (in integral form)
\begin{equation} \label{beta}
\beta =\frac{1}{{R_{h}^{2}}}\int \frac{dF_h}{|\kappa _h|}.
\end{equation}
Similarly, for the choice in Eq. (\ref{mod2}), the dimensionless parameter $\delta$ has the expression
\begin{equation} \label{gamma}
\delta =\frac{1}{R_{h}^{2}}\text{exp}\left[\int \frac{dF_h}{R_{h}^{2} |\kappa _h|}\right].
\end{equation} 
It should be noted that for the above two possible modifications of entropy and temperature, the FLT is valid for general (non-trivial) LTB models provided the paramterers involved are given by the relations (\ref{beta}) and (\ref{gamma}) respectively.

\section{CONCLUSIONS}

The present work deals with thermodynamical analysis of the Universe bounded by an arbitrary horizon for the inhomogeneous LTB model. At first, we define the GHT on the horizon and then examine the validity of the UFLT for the following three possible choices of entropy and temperature on the horizon---
\begin{equation} \label{3cases}
\left.
\begin{array}{rclrcl}
S_{h}^{(m)} & = & S_{h},~~& T_{h}^{(m)} & = & \alpha T_{h}, \\
S_{h}^{(m)} & = & \beta S_{h},~~& T_{h}^{(m)} & = & T_{h}, \\
S_{h}^{(m)} & = & \delta S_{h},~~& T_{h}^{(m)} & = & \frac{1}{\delta} T_{h}.
\end{array}
\right\}
\end{equation}
For the first choice of temperature and entropy, the FLT has been shown to be valid (for any choice of $\alpha$) only for the de Sitter particular case of the model, i.e., the horizon satisfies the de Sitter condition, while the validity of the GSLT restricts the evolution of the horizon radius, i.e., the horizon radius must be an increasing function of the temporal as well as the radial coordinate. On the other hand, for the other two choices of temperature and entropy, the dimensionless parameters $\beta$ and $\delta$ [given in Eqs. (\ref{beta}) and (\ref{gamma})] are obtained only in integral form, demanding the validity of the FLT. It should be noted that if $T_h$ is taken to be $T_{h}^{(g)}$ in Eq. (\ref{mod1}), then $\beta$ turns out to be unity. Further, due to integral expressions for $\beta$ and $\gamma$, it is not possible to examine the validity of the GSLT for these two choices of temperature and entropy. Hence it should be kept in mind that the model need not be restricted to the de Sitter particular case, rather a general LTB model for the validity of the FLT for the last two choices in Eq. (\ref{3cases}). 

Moreover, the integrability condition of the total entropy variation Eq. (\ref{gge}) has been investigated in Appendix A. It has been found that the integrability condition is satisfied if $\alpha$ is constant, or $\alpha$ is proportional to the horizon radius, or $\alpha$ has an integral form [see Eq. ($A6$)], provided the particular de Sitter form of the model is chosen. 
Further, it has been shown recently \cite{Chakraborty4} that a FRW Universe bounded by an event horizon is a generalized Bekenstein-Hawking system but similar analysis for the present LTB model is not valid (or possible) even for an apparent horizon. Therefore the idea of Bekenstein system is no longer valid for an inhomogeneous perfect fluid LTB thermodynamical system.

\begin{acknowledgments}

One of the authors (S.C.) is thankful to IUCAA, Pune, India for their warm hospitality as a part of the work was done during a visit there. SC also acknowledges the UGC-DRS Programme in the Department of Mathematics, Jadavpur University. The author S.S. is thankful to UGC-BSR Programme of Jadavpur University for awarding research fellowship.

\end{acknowledgments}


\section*{Appendix A: Integrability Condition of Eq. (\ref{gge})}

From Eq. (\ref{gge}), we have
\begin{eqnarray}
\frac{\partial S_T}{\partial t} &=& \left(\frac{4\pi R_{h}^{3}\rho}{T_{h}^{(g)}}\right)\left\lbrace \frac{\dot{R}_h}{R_h}-(1+w)H \right\rbrace \nonumber \\
\frac{\partial S_T}{\partial r} &=& \left(\frac{4\pi R_{h}^{3}\rho}{T_{h}^{(g)}}\right)\left\lbrace (2+w)\frac{R_{h}^{'}}{R_h}-\frac{w'}{3w} \right\rbrace \nonumber .
\end{eqnarray}
So,
\begin{eqnarray}
\frac{\partial ^2 S_T}{\partial r \partial t} &=& 4\pi \frac{\partial}{\partial r}\left(\frac{R_{h}^{3}\rho}{T_{h}^{(g)}}\right)\left\lbrace \frac{\dot{R}_h}{R_h}-(1+w)H \right\rbrace +\left(\frac{4\pi R_{h}^{3}\rho}{T_{h}^{(g)}}\right) \Biggl\{\frac{{\dot{R}_h}{'}}{R_h}-\frac{\dot{R}_h R_{h}^{'}}{R_{h}^{2}} \nonumber \\
&-& (1+w)H'-w'H \Biggr\}, \nonumber \\
\frac{\partial ^2 S_T}{\partial t \partial r} &=& 4\pi \frac{\partial}{\partial t}\left(\frac{R_{h}^{3}\rho}{T_{h}^{(g)}}\right) \left\lbrace (2+w)\frac{R_{h}^{'}}{R_h}-\frac{w'}{3w} \right\rbrace +\left(\frac{4\pi R_{h}^{3}\rho}{T_{h}^{(g)}}\right)\Biggl\{ (2+w)\frac{{\dot{R}_h}{'}}{R_h} \nonumber \\
&-& (2+w)\frac{\dot{R}_h R_{h}^{'}}{R_{h}^{2}}+\dot{w}\frac{R_{h}^{'}}{R_h}-\frac{{\dot{w}}'}{3w}+\frac{w' \dot{w}}{w^2} \Biggr\} . \nonumber
\end{eqnarray}
As we have $1+w=0$ for the validity of the FLT, so for integrability $\left(\text{i.e.,}~\frac{\partial ^2 S_T}{\partial r \partial t}=\frac{\partial ^2 S_T}{\partial t \partial r}\right)$, we obtain
$$\frac{\partial}{\partial r}\left(\frac{R_{h}^{3}\rho}{T_{h}^{(g)}}\right)\frac{\dot{R}_h}{R_h}=\frac{\partial}{\partial t}\left(\frac{R_{h}^{3}\rho}{T_{h}^{(g)}}\right)\frac{R_{h}^{'}}{R_h}.  \eqno(A1)$$
Also, for $1+w=0$, the energy-momentum conservation relations (7) gives $\rho =\text{constant}$. Now using $T_{h}^{(g)}$ from Eq. (15), i.e., $T_{h}^{(g)}=\alpha \frac{|\kappa _h|}{2\pi}$, Eq. ($A1$) simplifies to
$$\frac{\partial}{\partial r}\left(\frac{1}{\kappa_h \alpha}\right)\dot{R}_h=\frac{\partial}{\partial t}\left(\frac{1}{\kappa_h \alpha}\right)R_{h}^{'}.  \eqno(A2)$$
Using Eq. (\ref{frd-fr}) in Eq. (11), the expression for the surface gravity becomes
$$\kappa _h=-4\pi R_h \rho +\frac{F}{2R_{h}^{2}}.  \eqno(A3)$$
Now using the field equations (\ref{efeltb}) and the condition (\ref{frd-fr}), we have
$$\left.
\begin{array}{rcl}
\frac{\partial \kappa _h}{\partial r} & = & -\frac{FR_{h}^{'}}{R_{h}^{3}}, \\
\frac{\partial \kappa _h}{\partial t} & = & -\frac{F \dot{R}_{h}}{R_{h}^{3}}.
\end{array}
\right\} \eqno(A4)$$
Hence the integrability condition ($A2$) now becomes
$$\frac{\partial \alpha}{\partial r}\dot{R}_{h}=\frac{\partial \alpha}{\partial t}R_{h}^{'},  \eqno(A5)$$
i.e.,
$$\left.
\begin{array}{rcl}
\alpha {'} & = & \mu (r,t)R_{h}^{'}, \\
\dot{\alpha} & = & \mu (r,t)\dot{R}_{h},
\end{array}
\right\} \eqno(A5a)$$

or equivalently,
$$\alpha =\int \mu (r,t)dR_h .  \eqno(A6)$$
Thus integrability condition is satisfied provided any one of the following conditions hold---\\
($a$) $\alpha$ is a constant,\\
($b$) $\alpha =\mu R_h$ when $\mu$ is a constant,\\
($c$) $\alpha$ is given by ($A6$) for any arbitrary $\mu$.

\end{document}